# The discovery of a high redshift type 2 QSO

O. Almaini[1], B.J. Boyle[2], R.E. Griffiths[3], T. Shanks[1], G.C. Stewart[4] and I. Georgantopoulos[4]

[1] *Department of Physics, University of Durham, South Road, Durham, DH1 3LE, UK.*
[2] *Royal Greenwich Observatory, Madingley Road, Cambridge, CB3 OEZ, UK.*
[3] *Johns Hopkins University, Homewood Campus, Baltimore MA 21260, USA.*
[4] *Department of Physics, University of Leicester, University of Leicester, LE1 7RH, UK.*

29 August 1995

**ABSTRACT**

We report the discovery of a high redshift, narrow emission-line galaxy identified in the optical follow-up of deep *ROSAT* fields. The object has a redshift of $z = 2.35$ and its narrow emission lines together with its high optical and X-ray luminosity imply that this is a rare example of a type 2 QSO. The intrinsic X-ray absorption is either very low or we are observing scattered flux which does not come directly from the nucleus. The X-ray spectrum of this object is harder than that of normal QSOs, and it is possible that a hitherto unidentified population of similar objects at fainter X-ray fluxes could account for the missing hard component of the X-ray background.

## INTRODUCTION

We are currently conducting a major observational programme to understand the nature of the X-ray background. Using optical spectroscopy we attempt to identify the optical counterparts to faint X-ray sources detected on deep (30 – 50 ksec) *ROSAT* PSPC exposures. So far we have identified over 150 QSOs from 7 *ROSAT* fields and have shown that QSOs directly make up at least 30% of the X-ray background (XRB) (Shanks et al 1991) and could comprise up to 50% of the XRB at 1 keV (Boyle et al 1994, Georgantopoulos et al 1993). The origin of the remainder of this background radiation is still unresolved but it is unlikely to be due to ordinary QSOs. These show relatively steep X-ray spectra with indices of $\Gamma = 2.2 \pm 0.1$ while the extragalactic X-ray background from 1-10 keV has a flatter power law index of $\Gamma = 1.4$ (Gendreau et al 1995). This suggests that we need a new, faint source population with a flatter X-ray spectrum to account for the remainder of the X-ray background.

Recent work by Roche et al (1995) has shown that faint B<23 galaxies may contribute up to 30% of the X-ray background signal at 1 keV but the implied X-ray luminosities are ~ 10-100 times greater than for similar galaxies locally. Further work by Almaini et al (in preparation) has shown that these galaxies have flatter X-ray spectra than QSOs, more consistent with the residual background radiation, but the nature of the emission mechanism in these galaxies is still unresolved. One possibility is that some of these galaxes harbour obscured AGN.

In this paper we present the discovery of an unusual narrow emission line object (RXJ13434+0001) detected during the optical spectroscopic follow up of two new deep *ROSAT* fields. Optical and X-ray observations are presented in Section 2 followed by a discussion on the nature of this object and the possible implications for our understanding of the X-ray background.

## 2 OBSERVATIONS

### 2.1 Optical Identification

We recently obtained 2 new 50 ksec *ROSAT* exposures centred on the QSO survey fields F855 and F864 (Boyle et al 1991) and in March 1994 we carried out optical spectroscopic follow-up using Autofib and the RGO Spectrograph at the Anglo-Australian Telescope. We obtained 12Å resolution spectra for the optical counterparts to ~ 80 X-ray sources detected in each field. Figure 1 shows the optical spectrum of a narrow emission-line object identified on the F864 field. The optical position of this object is $13^h 43^m 29.2^s + 00°01'33''$ (J2000). AAT plates identify RXJ13434+0001 as a star-like object at $B = 22$, only $3''$ from the X-ray source position and well within the $\simeq 15''$ $2\sigma$ X-ray error box. Identification of the lines as Ly$\alpha$ and CIV$\lambda$1549 give a redshift of $z = 2.347 \pm 0.006$, implying an absolute magnitude of $M_B = -23.7$ ($H_0 = 50 \, \mathrm{km\,s^{-1} Mpc^{-1}}$, $q_0 = 0.5$).

Taking into account the instrumental resolution, the full width at half maxima for the Ly$\alpha$ and CIV$\lambda$1549 emis-sion lines are $900 \, \text{km s}^{-1}$ and $< 700 \, \text{km s}^{-1}$ respectively. Based on the widths and strengths of these lines (the presence of highly ionized CIV$\lambda$1549 is inconsistent with a starburst galaxy), we classify RXJ13434+0001 as a type 2 AGN. Given its optical and X-ray luminosity, RXJ13434+0001 is therefore a rare example of a type 2 QSO (Stocke et al 1982, Elizalde& Steiner 1994) as required by unified schemes for AGN (Antonucci et al 1985).

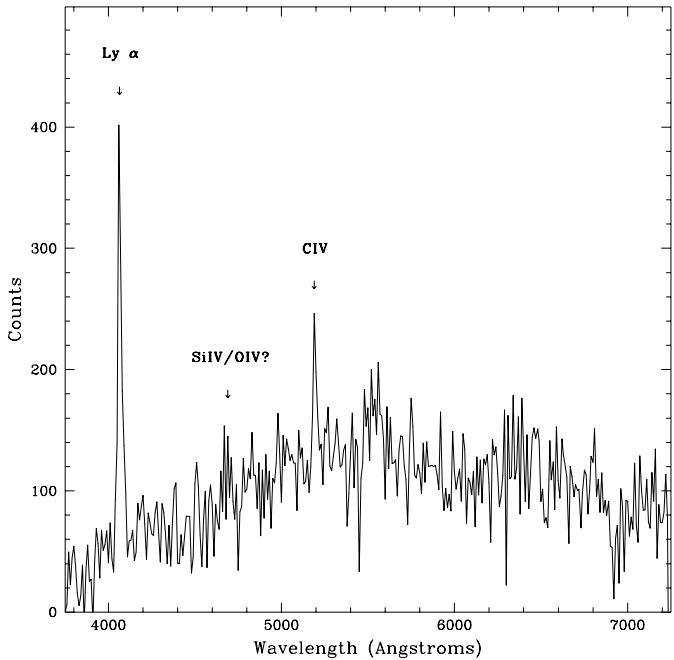

**Figure 1.** Low resolution (12Å) optical spectrum for RXJ13434+0001. The prominent emission lines are marked.

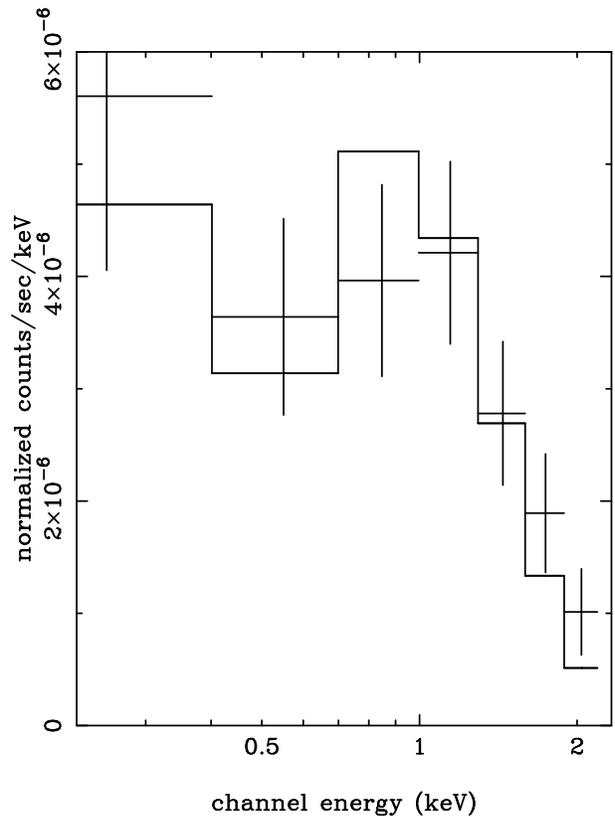

**Figure 2.** The observed 0.1 − 2 keV X-ray spectrum with the best fitting power law model ($\Gamma = 1.56$) modified by galactic absorption.

### 2.2 X-ray Observations

RXJ13434+0001 was detected as a 15$\sigma$ source with a 0.1 − 2 keV X-ray flux of $4.5 \times 10^{-14} \, \text{erg s}^{-1} \text{cm}^{-2}$. The best fitting power law spectrum gives a 0.4−6.7 keV rest-frame luminosity of $1.9 \times 10^{45} \, \text{erg s}^{-1}$. Fixing the absorption at a galactic column of $N_H = 2.6 \times 10^{20} \, \text{atom cm}^{-2}$, a power law model gave a good fit to the X-ray spectrum ($\chi^2_{\text{red}} = 1.07$) with a photon index of $\Gamma = 1.56$ and a 95% upper limit of $\Gamma < 1.98$ with one interesting parameter, significantly flatter than the mean photon index $\Gamma = 2.2 \pm 0.1$ obtained for other QSOs in our sample by the same method. The raw channel spectrum is shown on Figure 2 with the best fitting model.

### 3 DISCUSSION

At rest-frame energies of 0.4 − 6.7 keV, most local type 2 Seyferts show strong photo-electric absorption columns of $N_H = 10^{22} − 10^{24} \, \text{atom cm}^{-2}$ (Mulchaey et al 1985). For RXJ13434+0001 the best-fit cold absorber model gives zero for the intrinsic (i.e. rest-frame) absorption, but a column of $N_H = 6 \times 10^{21} \, \text{atom cm}^{-2}$ is permitted within the 99% confidence contours (see Figure 3). We conclude that an absorbing column could be present at the lower end of the range seen in Seyfert 2 galaxies. The dust associated with a column of $N_H \sim 3 \times 10^{21} \, \text{atom cm}^{-2}$ would be enough to obscure the broad line region in the ultra-violet band assuming standard gas to dust ratios, producing at least 5 magnitudes of UV extinction. However, the UV continuum is somewhat brighter than we might expect from this model. Fitting the standard hypothetical power law between the UV and soft X-ray spectrum at 2 keV we obtain $\alpha_{ox} = 1.14$ compared to $\alpha_{ox} = 1.5 \pm 0.1$ for normal $z = 2$ QSOs (eg. from Boyle et al 1994), which suggests some UV extinction but only $\sim 2.5$ magnitudes. This could be explained if the UV continuum had some contribution from a bright host galaxy. Alternatively, if RXJ13434+0001 genuinely has an intrinsic $N_H < 3 \times 10^{21} \, \text{atom cm}^{-2}$ the broad line region cannot be obscured and this would raise the intriguing possibility that we may have discovered a 'naked' type 2 nucleus. Another explanation is that the observed X-rays are due to scattered flux (Fabian et al 1994) and the absorbing column is actually very large ($> 10^{24} \, \text{atom cm}^{-2}$). If this is the case we would expect an iron K fluorescence line in the X-ray spectrum at a redshifted energy of 2 keV, but we require better data before we can determine whether this line is present.

We consider it unlikely that RXJ13434+0001 belongs to the rare class of narrow-line Seyfert 1 galaxies (Brandt et al 1994). These objects generally show steep X-ray spectra with $\Gamma > 2.5$, are far less luminous than RXJ13434+0001 and they also tend to exhibit rapid X-ray variability. Our object has a significantly flatter spectrum, $\Gamma = 1.56$, and shows no evidence for variability in the limited data we have. Indeed, the steep spectrum of these sources is actually invoked as the reason for the narrow lines since a steep X-ray spectrum might interfere with the formation and confinement of broad line clouds. This clearly cannot be the case for our object.

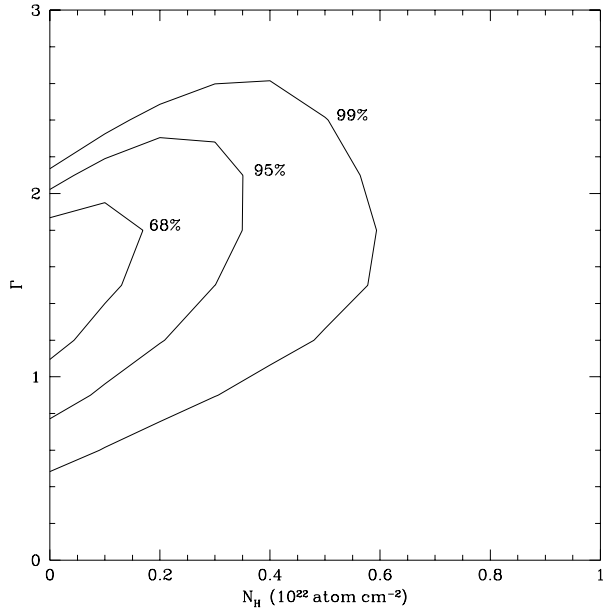

**Figure 3.** Confidence contours in the rest frame column density ($N_H$) v. photon index ($\Gamma$) plane for the X-ray spectral fits to RXJ13434+0001. The contours represent the 68%, 95% and 99% confidence regions with two interesting parameters.

It is difficult to make claims on the basis of one new object, but nevertheless we note that the discovery of a population of type 2 QSOs with hard X-ray spectra would have important implications for the origin of the X-ray background. A current problem is that the spectra of QSOs are different to that of the residual background radiation. *ASCA* observations (Gendreau et al 1995) have shown that the extragalactic XRB from $1 - 10\,\mathrm{keV}$ has a flat power-law slope of $\Gamma = 1.4$ while type 1 QSOs have a much steeper slope, $\Gamma \simeq 2$, suggesting that a new population with a flatter X-ray spectrum is required to account for the remainder. Recent modelling (Comastri et al 1995, Madau et al 1994) has shown that the entire $5 - 100\mathrm{keV}$ X-ray background and $\sim 70\%$ of the *ROSAT* band may be explained in the context of unified schemes in which we have a range of obscured and unobscured AGN. An important prediction of this model is the existence of a large population of previously undetected obscured QSOs, many of which should be detectable at faint fluxes even in the *ROSAT* band. One explanation for the new population of low redshift X-ray galaxies emerging from *ROSAT* surveys (eg. Roche et al 1995, Boyle et al 1995) is that some of these galaxies contain hidden AGN. The QSO RXJ13434 + 0001 may therefore represent a particularly bright high redshift counterpart to these objects. If an obscured QSO population does exist at high redshift they could make a significant contribution to the X-ray background but still remain too faint to observe any optical counterparts on photographic plates to the limits probed by our survey (B<23). From an X-ray absorbing column typical of Seyfert 2 galaxies we would expect at least 5 magnitudes of extinction in the rest frame UV which corresponds to the observed optical waveband. It is therefore not surprising that such objects have not been detected before. Uncovering this obscured population may therefore be the key to resolving the origin of the X-ray background.

## ACKNOWLEDGMENTS

OA was funded by an SERC/PPARC studentship. BJB was partly supported by a Royal Society University Research Fellowship during part of this work. We are indebted to Christine Done for her help and enthusiasm and to Andy Fabian for useful discussions. We are also grateful to the Director and staff of the Anglo-Australian observatory and the *ROSAT* team for the high quality of support received throughout our observing programme.